\documentclass[prb,twocolumn,showpacs,preprintnumbers,amsmath,amssymb]{revtex4}

\usepackage{graphicx}
\usepackage{dcolumn}
\usepackage{bm}

\begin{document}

\title{Conduction Channels of One-Atom Zinc Contacts}

\author {M. H\"afner,$^1$ P. Konrad,$^2$ F. Pauly,$^1$ J.C. Cuevas,$^1$
and E. Scheer$^2$}

\affiliation{$^1$Institut f\"ur Theoretische Festk\"orperphysik,
Universit\"at Karlsruhe, 76128 Karlsruhe, Germany\\
$^2$Fachbereich Physik, Universit\"at Konstanz, 78457 Konstanz, Germany}

\date{\today}

\begin{abstract}
 We have determined the transmission coefficients of atomic-sized Zn contacts
 using a new type of breakjunction which contains a whisker as a central bridge.
 We find that in the last conductance plateau the transport is unexpectedly
 dominated by a well-transmitting single conduction channel. We explain the
 experimental findings with the help of a tight-binding model which shows that
 in an one-atom Zn contact the current proceeds through the $4s$ and $4p$ orbitals
 of the central atom.
\end{abstract}


\maketitle

The appearance of experimental techniques such as the scanning
tunneling microscope and breakjunctions has allowed to explore the
electronic transport at the atomic scale~\cite{Agrait2003}. With
these techniques it is possible to gently break a metallic contact
and thus form conducting nanowires. During the last stages of the
pulling a neck-shaped wire connects the two electrodes, the
diameter of which is reduced to a single atom upon further
stretching. The conductance of these contacts can be described by
the Landauer formula: $G = G_0 \sum_i \tau_i$, where the sum runs
over all the available conduction channels, $\tau_i$ is the
transmission for the $i^{th}$ channel and $G_0 =2e^2/h$ is the
quantum of conductance. As it was shown in
Ref.~[\onlinecite{Scheer1997}], the set of transmission
coefficients is amenable to measurement in the case of
superconducting materials. Using this possibility it has been
established that the number of channels in an one-atom contact is
determined by the valence of the metal, and the transmission of
each channel is fixed by the local atomic
environment~\cite{Cuevas1998a,Scheer1998,Cuevas1998b}. Thus for
instance, an one-contact of a monovalent metal such as Au sustains
a single channel, while for $sp$-like metals such as Al or Pb one
finds three channels due to the contribution of the $p$ orbitals,
and in a transition metal such as Nb the contribution of the $d$
orbitals leads to five channels. Up to now, these attractive ideas
have only been tested in four materials (Au, Al, Pb, and Nb) due
to the need of superconductivity for the channel analysis. In this
sense, it would be highly desirable to investigate other groups of
metals. An interesting possibility is the analysis of the divalent
metals of the IIB group of the periodic table such as Zn. The
electronic structure of a Zn atom is $[\mbox{Ar}]3d^{10} 4s^2$,
i.e. the outermost $s$ orbital is full with two electrons. As a
solid, Zn is a conductor due to the overlap between the $4s$ and
the $4p$ bands. Therefore, for one-atom contacts one expects Zn to
be an intermediate case between the noble metals and Al. The goal
of this work is to elucidate what determines the
conduction channels of one-atom Zn contacts.

\begin{figure}[b]
\begin{center}
\includegraphics[width=0.9\columnwidth,clip]{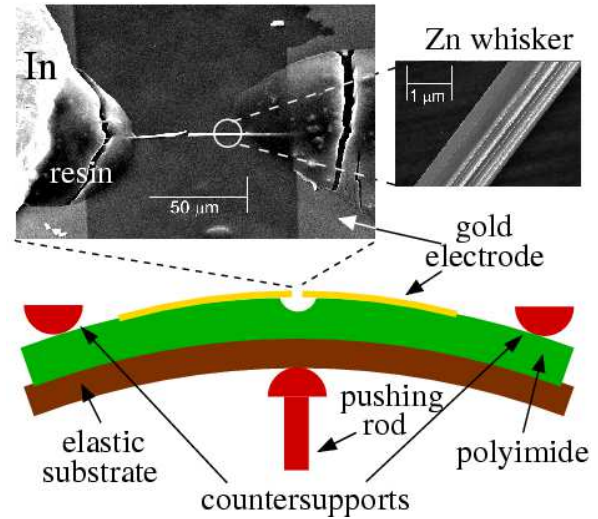}
\caption{\label{setup} (Color online) Three-point bending mechanism. The
pushing rod bends the bronze substrate. The optical micrograph shows a
whisker MCB after measurement. Right inset: blow up of a typical
Zn whisker with a diameter of 1.7~$\mu$m.}
\end{center}
\end{figure}

Traditionally, oxidation problems and the gap
anisotropy~\cite{Cleavelin1974} have prevented the formation of
reproducible superconducting contacts with Zn. In order to
circumvent these problems we have developed a variation of the
mechanically controllable breakjunction (MCB) technique which
comprises a Zn whisker as central part. This new technique allows
us to fabricate reproducibly one-atom Zn contacts with
well-characterized lattice properties of the electrodes. Using the
superconducting current-voltage $(IV)$ characteristics~\cite{Scheer1997}
we have reliably extracted the transmission coefficients. We find
that in the last plateau, where the conductance is around $0.8~G_0$
(Refs.~[\onlinecite{Yanson2001,amg2004}]), the transport is largely
dominated by a single channel. In order to understand these findings
we have performed tight-binding calculations of the conductance of
one-atom Zn contacts. Our theoretical results show that the
transport takes place through the $4s$ and $4p$ bands of Zn. In
agreement with the experiment, we obtain for the one-atom case a
conductance between $0.8$ and $1.0~G_0$, and it is dominated by a
single channel which is a combination of the $s$ orbital and the
$p$ orbital along the transport direction of the central atom.

\begin{figure}[t]
\begin{center}
\includegraphics[width=\columnwidth,clip]{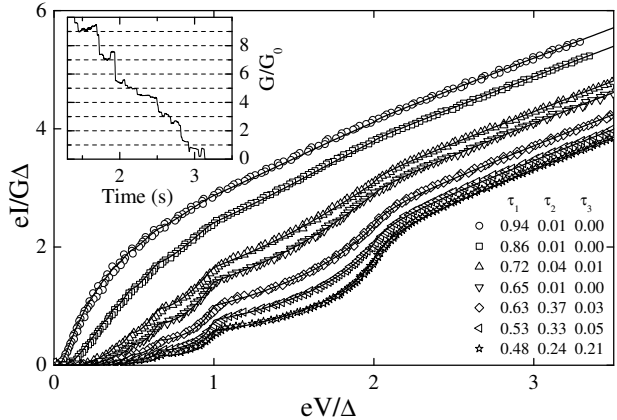}
\caption{\label{ivs} Current-voltage characteristics of several
one-atom contacts of Zn arranged with a whisker MCB at $T =
0.27$~K (symbols) and best fits to MAR
theory~\cite{Averin1995,Cuevas1996}. The transmission coefficients
obtained from the fit are indicated in the figure. Inset: typical
opening curve recorded at $T = 1.5$~K. }
\end{center}
\end{figure}

Our first attempts to investigate Zn contacts were carried out
with both ``notched-wire" breakjunctions~\cite{Muller1992} and
thin-film breakjunctions~\cite{Ruitenbeek1996}. However, the
granular structure of the evaporated films and the intrinsic gap
anisotropy of Zn prevented the observation of reproducible
superconducting $IV$s, which are necessary to obtain the channel
transmissions. To solve these problems we have prepared MCBs
consisting of Zn whiskers glued on top of a pre-patterned flexible
substrate (see Fig.~\ref{setup}). Whiskers are thin single
crystalline needles, which in the case of Zn grow with a
probability of 67\% along the $a$-axis by the so-called pressure
method \cite{Schulz1987}. Following Ref.~[\onlinecite{Schulz1987}]
we have produced whiskers by electroplating $8~{\rm to}~10~\mu$m
thick layers of 99.99\% pure Zn onto stainless steel
substrates~\cite{note1}. We apply a uniaxial pressure of 50
N/mm$^2$ onto a package of 15 substrates, polish its edges
carefully and store it at $80~^{\circ}$C and 200~mbar helium
atmosphere. The whiskers start to grow from the edges of the
package. The growth saturates after a period of 3 to 6 months,
giving rise to whiskers of diameter ranging from 0.5 to 2 $\mu$m
and length of the order of 0.5 to 1~mm (see inset of
Fig.~\ref{setup}). To contact the whiskers we prepare bronze
substrates of size 3$\times$18~mm$^2$ covered with an insulating
layer of polyimide and 70~nm thick gold electrodes separated by
100~$\mu$m. Then, individual whiskers are deposited onto the
substrate and mechanically contacted by two small dots of epoxy
resin. The electrical contact between the whiskers and the
electrodes is obtained by gluing two small pieces of In onto the
whiskers and the gold pads. With this method we obtain whisker
breakjunctions with resistances of 10 to 100~$\Omega$ at room
temperature. Finally, we mount them onto a three point bending
mechanism~\cite{Muller1992} (see Fig.~\ref{setup}) thermally
anchored to the base temperature pot ($\approx 260~$mK) of a
$^3{\rm He}$ cryostat.

\begin{figure}[t]
\begin{center}
\includegraphics[width=0.8\columnwidth,clip]{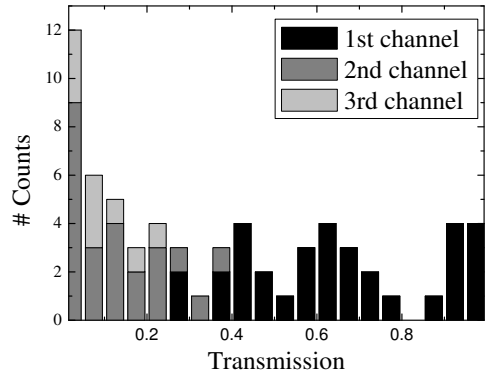}
\caption{\label{histogram} Experimental histogram of the channel
transmissions $\tau_i$ obtained in the last conductance plateau of
a Zn-whisker MCB. We count all those contacts as belonging to the
last plateau that have been recorded when the conductance once
dropped to or below $G_0$. The channels are labeled as 1st, 2nd,
... according to their $\tau_i^{\prime}s$.}
\end{center}
\end{figure}

Breaking the contact, the conductance evolves as usual in a
step-like manner. An example of an opening curve is shown in the
inset of Fig.~\ref{ivs}, where one sees the typical behavior for
Zn atomic-sized contacts with a conductance in the last plateau of
the order of $0.8~G_0$. The stability of the setup allows us to
halt at any point along the breaking to record the $IV$s. In
particular, to determine the superconducting gap the whisker of
Fig.~\ref{setup} was broken to form a tunnel contact. In this
limit a well-pronounced gap is visible in the $IV$s, and a value
of $\Delta = 160$~$\mu$eV was deduced. This value is close to the
expected value for the $x$-direction in real space, corresponding
to the crystallographic [110] direction, the direction with the
largest gap. Although this suggests that the whisker was grown in
this direction, we cannot exclude that this rather high value is
due to the proximity effect with the In contacts ($2\Delta_{In}
\simeq 1.05$~m$e$V). In Fig.~\ref{ivs} we show the superconducting
$IV$s obtained at 0.27~K for seven different contacts.
Together with the data points we also show the best fit obtained
with the multiple Andreev reflection (MAR)
theory~\cite{Averin1995,Cuevas1996} at zero temperature. The
details of the fit procedure are explained in
Ref.~[\onlinecite{Scheer1997}]. The excellent agreement with the
MAR theory allows us to determine the whole set of transmission
coefficients $\{ \tau_i \}$ with a precision of a few percent
(depending on the value of $\tau_i$) for up to 4
channels. To investigate the conduction channels of the smallest
contacts (presumably one-atom contacts), we have collected the
transmission coefficients obtained in the last plateau of 33
contact realizations. The results are shown in the histogram of
Fig.~\ref{histogram}. From this plot we draw two important
conclusions: (i) the conductance in the last plateau is largely
dominated by a single channel with a high transparency, and (ii)
depending on the contact geometry a second and third channels with
low transmissions can also contribute to the transport.

In order to understand the experimental findings we have performed
conductance calculations following the method described in
Refs.~[\onlinecite{Cuevas1998a,Cuevas1998b}].
Briefly, in this approach the electronic structure of
the atomic contacts is described in terms of a tight-binding
Hamiltonian written in an orthogonal basis. The different on-site
energies and hopping elements are taken from the bulk parametrization
of Ref.~[\onlinecite{Papa1986}], which accurately reproduces the
band structure of bulk materials. Our basis is formed by 9
atomic orbitals: $3d,4s,4p$, which give rise to the main bands
around the Fermi energy, and hopping elements up to third neighbors
are taken into account. We have imposed local charge neutrality
in all the atoms of the constriction by means of a self-consistent
variation of the on-site energies. The leads are described with bulk
atoms. The transmission of the contacts is calculated using Green
function techniques, which allow us to express the set of transmission
coefficients $\{ \tau_i \}$ in terms of the microscopic parameters
of the atomic contacts.

\begin{figure}[t]
\begin{center}
\includegraphics[width=\columnwidth,clip]{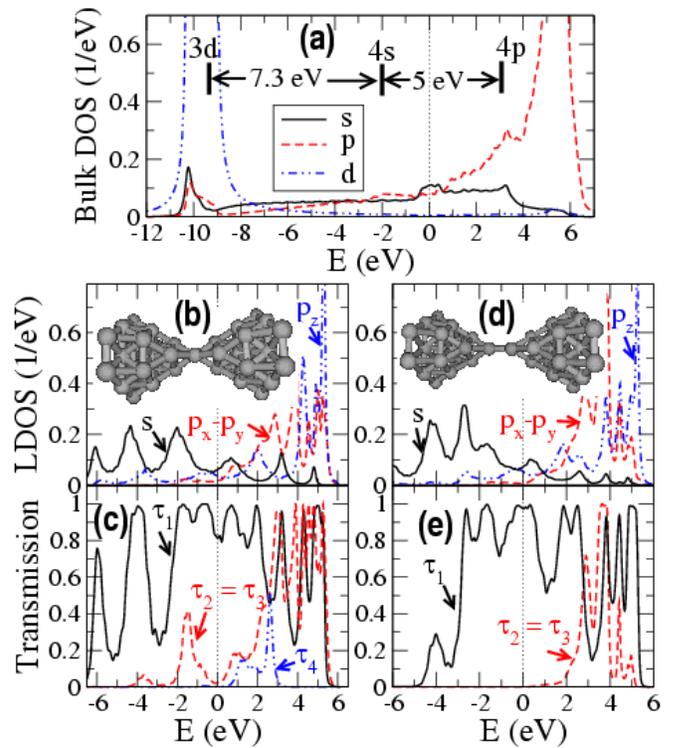}
\caption{\label{theory1} (Color online) (a) Bulk density of states 
of Zn hcp projected onto the $s$, $p$ and $d$ orbitals as a function of
energy. The labels indicate the value of the on-site energies.
Inset of panel (b): contact along the [001] direction, with a
single central atom and containing 45 atoms. The distance between
the different atoms is the bulk distance. (b-c) Local DOS at the
central atom projected onto the $s$ and $p$ orbitals and channel
transmissions as a function of energy for the contact of the inset
of panel (b). Notice that the channel $p_x$-$p_y$ is two-fold
degenerate. Inset of panel (d): the same contact as in panel (b),
but with a dimer in the narrowest part. The distance between the
central atoms is the bulk nearest-neighbor distance. (d-e) The
same as in panels (b-c) for the dimer geometry. The Fermi energy
is set to zero and indicated with vertical lines.}
\end{center}
\end{figure}

It is instructive to first discuss the bulk
density of states (DOS) of Zn in its hcp structure, which is
shown in Fig.~\ref{theory1}(a). Notice that the $d$ band is rather
narrow, it lies $\approx 9$~eV below the Fermi energy and it is
practically full. It is then obvious that the $s$ and $p$
bands will play the main role in the conduction. Since there is an
uncertainty in the growth direction of the whiskers, we have
studied the conductance of geometries with different
crystallographic orientations. In the inset of
Fig.~\ref{theory1}(b) we show an example of an one-atom contact
along the [001] direction ($c$-axis). The geometry is constructed
starting with a single atom and choosing the nearest-neighbors in
the next layers. In Fig.~\ref{theory1}(b-c) one can see for this
geometry the local DOS at the central atom and the transmission of the
individual channels $\tau_i$ as a function of energy. In the local
DOS we see that the $p_z$ level ($z$ is the transport direction)
is shifted to lower energies due to its better coupling to the
leads as compared with $p_x$ and $p_y$, which remain degenerate in
this ideal geometry. This fact implies that the orbital $p_z$
plays a more important role in the transport. In the relevant energy
range the $d$ band has a very tiny local DOS and it has therefore not been
depicted. In Fig.~\ref{theory1}(c) we see that at the Fermi energy
the total transmission is 0.86 and it is dominated by a single
channel with $\tau_1 = 0.81$. The second and third channel are
degenerate and their transmission at the Fermi energy is $\tau_2
= \tau_3 = 0.025$, while the fourth gives a negligible
contribution $\tau_4 = 0.001$. To understand the origin of these
conduction channels we have analyzed the character of the
eigenfunctions of the transmission matrix by looking at their
weights in the different orbitals of the central atom. This
analysis reveals that the dominant channel is basically a
symmetric combination of the $s$ and $p_z$ orbitals of the central
atom. The second nd third channels are mainly due to the
contribution of the $p_x$ and $p_y$ orbitals. The degeneracy of
these two channels is a consequence of the symmetry of this ideal
geometry and reflects the degeneracy of the local DOS, see
Fig.~\ref{theory1}(b). Their transmission is rather low because
the transport takes place through the tails of the $p_x$ and $p_y$
bands. The antisymmetric combination of $s$ and $p_z$ forms a
channel of negligible transmission due to the fact that
this combination is orthogonal to the states of the leads. So in
short, the nature of the channels in this one-atom
Zn contact is similar to the Al case~\cite{Cuevas1998a}. The main
difference is that Zn has one valence electron less than
Al. Thus, the Fermi energy is lower and consequently
lies further away from the center of the $p$ bands, resulting in a
lower transmission of the $p_x$-$p_y$ channels.

As explained above, the second and third channels have a low
transmission due to the fact that the transport takes place
through the tails of the $p_x$ and $p_y$ bands. This suggests that
a reduction of the coupling to these orbitals could result in a
negligible contribution of these channels. To test this idea we
have analyzed one-atom contacts with a dimer in the narrowest part
of the contact, see inset of Fig.~\ref{theory1}(d). This type of
geometry has been frequently observed in molecular dynamics
simulations of atomic contacts of Al~(Ref.~[\onlinecite{Jelinek2003}])
and Au~(Ref.~[\onlinecite{Dreher2004}]). In Fig.~\ref{theory1}(d-e) the
local DOS in one of the central atoms and the transmission through this
dimer contact are shown. The total transmission at the Fermi
energy is 0.97, and as suspected, it is completely dominated by a
single channel ($\tau_2 = 4\times 10^{-4}$), while the character
of the channels is the same as in the case analyzed above. Thus,
by changing the contact geometry from a short contact with a single
atom in the constriction to a long contact in a dimer
configuration the conductance may increase by about 0.1 $G_0$.
Opening traces that support this prediction are reported in the
literature~\cite{Yanson2001,amg2004} and have also been found in
the present experiment (see inset of Fig.\ref{ivs}).

\begin{figure}[t]
\begin{center}
\includegraphics[width=0.9\columnwidth,clip]{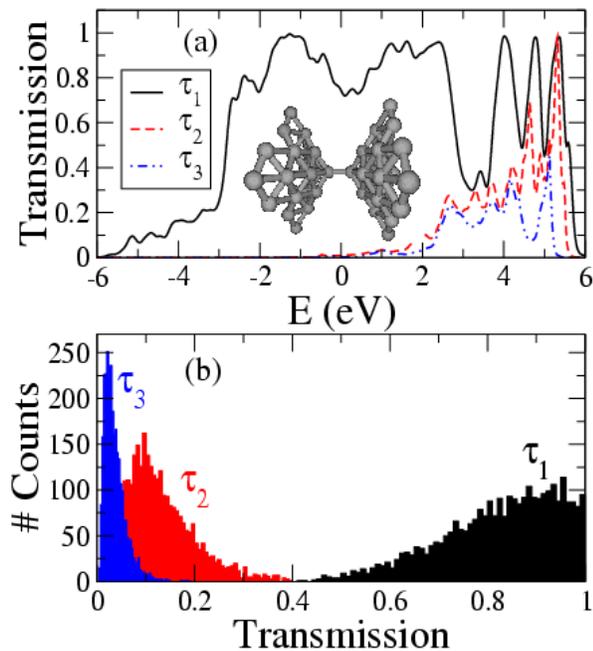}
\caption{\label{theory2} (Color online) (a) Channel transmissions as a
function of energy for the contact of the inset, which is grown along the [110]
direction, has a dimer in the middle and contains 60 atoms. The Fermi energy
is set to zero. (b) Histograms of the transmission coefficients for 3000
disorder realizations (see text) of the one-atom contact of the inset of
Fig.~\ref{theory1}(b). }
\end{center}
\end{figure}

Let us now discuss the results for contacts along the [110]
direction ($a$-axis). An example of a contact geometry with a
dimer configuration and its corresponding transmission is shown in
Fig.~\ref{theory2}. As in the case of the [001] direction, the
total transmission ($\tau_{tot} = 0.77$) is clearly dominated by a
single channel ($\tau_1 = 0.75$), which is formed by a symmetric
combination of the $s$ and $p_x$ orbitals of the central atoms
($x$ is now the transport direction). The second channel is on the
border of what is measurable ($\tau_2 = 0.01$), while the
transmission of the third one is $\tau_3 = 0.004$. Notice that in
this case these two channels are not degenerate. We find that the
transmission for contacts along the [110] direction is slightly
lower than for the [001] one, which we attribute to the larger
apex angle in the first direction, which produces less adiabatic
contacts. These findings are in agreement with the observation of
Ref.~[\onlinecite{amg2004}] where a splitting of the first peak of the
histogram into two sub-peaks with 0.7 and 0.9~$G_0$ was reported.

For all the geometries analyzed in this work we have checked that
the results do not change qualitatively with the number of atoms
in the constriction region. We have also studied the influence of
disorder, which we simulate by changing randomly the positions of
the atoms in the contact region with a very large maximum
amplitude of 20\% of the nearest-neighbor distance. The hoppings
are then computed using the scaling laws proposed in
Ref.~[\onlinecite{Papa1986}]. In Fig.~\ref{theory2}(b) we show a
histogram of the individual transmissions for 3000 realizations of
contacts along the [001] direction with a single central atom. As
it can be seen, the transmission is still dominated by a single
channel, and depending on the local environment of the central
atom a second and even a third channel can have a measurable
contribution, in agreement with the experimental results (see
Fig.~\ref{histogram}). Similar histograms for the dimer geometries
show that $\tau_2$ and $\tau_3$ typically lie below the measurement
threshold.

In summary, we have presented an experimental and theoretical
study of the conduction channels in Zn atomic junctions. We have
shown that, although Zn is a divalent metal, the conductance of
one-atom contacts is dominated by a single well-transmitting
channel, which we have traced back to the symmetry of the valence
orbitals ($s$ and $p$) of the central atom. Our results constitute
a new illustration of how the electronic structure of an atom
determines the conductance of the circuit in which it is embedded.

This work has been financed by the SFB~195, SFB~513, the Alfried
Krupp von Bohlen und Halbach-Stiftung, the Landesstiftung
Baden-W\"urttemberg and the Helmholtz Gemeinschaft (contract VH-NG-029).



\begin{thebibliography}{}
\bibitem{note1}
A detailed description of the fabrication procedure will be published elsewhere.
\end{thebibliography}
\end{document}